# STATISTICAL GEOMETRY OF GRAVITATION

Javier Bootello
jbootello@ciccp.es

ABSTRACT: General Relativity explains with precision the anomalous advance of Mercury´s perihelion, discovered by Le Verrier in 1859. Otherwise, diverse post-Newtonian proposals trying to solve this anomaly, introduce mathematical potentials focused on a finite propagation speed.

This paper tries to set some properties that should have any hypothetical gravitational potential suitable with material objects inside a physical universe. If a propagation speed is admitted, this assumption must link its origin, the continuous update, the potential´s trajectory, the retarded action and impulse mechanism "transit action" in the final target.

This gravitational potential, tries to explain the anomalous shift of the perihelion of Mercury. Otherwise, applied as a potential of force added to a mechanical momentum, could give a partial explanation of the anomalous acceleration of Pioneer 10/11. Anyway, it is an hypothetical conjecture without any proof of its physical reality.

## 1. - Gravitational Geometry. Mercury

In a flat space-time based on new post–Newtonian physics, any gravitational potential candidate to solve the anomalous precession of Mercury, must be related with a finite propagation speed and also with the relative radial velocity of the attracted object.

Laplace proposed in 1805 a gravitational force maintaining the inverse square of the distance law, adding a complementary term related with both speeds. This complementary force, was in the same direction that the vector-speed of the target. Also Riemann (1886) worked with a velocity–dependent gravitational force with the following expression:

$$F = G \cdot \frac{m_1 \cdot m_2}{r^2} \left( 1 - \frac{1}{c^2} \left( \frac{d\check{r}_1}{dt} - \frac{d\check{r}_2}{dt} \right)^2 \right)$$

In the hypothetical existence of a gravitational potential in the real universe of material objects, it seems reasonable to think that it could depend on the ratio between propagation speed (c) and that of the target. The gradient of a potential that moves forward with certain speed, will take a time to reach the target and also a very reduced time to cross it and produce a "transit action" in its mass. During that time, the attracted object, perceives the real existence of this potential and reacts by means of changing position, speed and trajectory. The impulse trasmitted by the potential to the target, becomes a balanced action to adapt itself to the force coming from the outside world.

Although it is considered unknown the nature of the iteration between the potential and the attracted body, if existing, it should be an electrodynamic one. If a potential and a target -with a speed (Vr)-, are moving with the same direction, the transit time between them will be larger if both have the same forward way, and will decrease if they are moving in the opposite one; the hypothetical iteration ("transit action") cannot be the same in each situation. The larger or reduced transit time between target and potential, is proportional to c / (c ±Vr), statement that gives us the relative increase or reduction ratio related with the transit time within the target in a rest position or a perfect orthogonal movement.

Be $t_1$ the transit time a potential uses to cross through an object. If the target moves in the same forward direction as the potential, the new time $t_2$ will be larger than $t_1$ and will have the following ratio:   $t_2 \cdot c = t_2 \cdot V_r + t_1 \cdot c$

$$\frac{t_2}{t_1} = \frac{c}{c - V_r} = \frac{1}{1 - V_r/c}$$

The resultant new force, will be however proportional to the square of this ratio as being electrodynamic Weber type force. Also, we must consider that energy potentials are always proportional to the square of the time. Therefore, a different behaviour from the target would be expected, if it moves forward or against the potential. However, this complementary "transit action" to the classic gravitational potential, would not be very significant since it is in proportion to $[c / (c \pm V_r)]^2$ that is in order of $10^{-9}$ in Mercury´s orbit. The radial speed of a planet is directly related with the eccentricity of its orbit and in Mercury the effect could be appreciate in spite of its reduced magnitude; if this new acceleration exist, it would perturb the orbit producing the advance of its perihelion. The modified gravitational attraction would have a potential with the following expression:

$$U(r, V_r) = - \frac{G M_\odot}{r} \cdot \left(\frac{t_2}{t_1}\right)^2 = - \frac{G M_\odot}{r} \left(\frac{1}{1 - V_r/c}\right)^2$$

The potential $U(r, V_r)$ is then similar to the newtonian with an added perturbation factor related with the square of the transit time ratio.

If we accept this interference and general formulation, it implies to consider that gravitation has a finite propagation speed and action at distance, should be a quantum electrodynamic iteration within the target body. It is necessary not to forget that although this statement could be theoretically admitted, we don't have any real or experimental confirmation of it.

A mathematical potential similar to $U(r, V_r)$, was studied by Gerber (1898) although it was based on a very different physical interpretation. We can obtain the value of the associated potential force, applying Lagrangian equations; however they give an appropriate mathematical result, its physical meaning admits alternatives. The numeric value of the apparent force is:

$$F = \frac{d}{dt}\left(\frac{\partial U}{\partial \dot r}\right) - \frac{\partial U}{\partial r} \quad ; \quad \dot r = V_r$$

obtaining from the potential $U(r, V_r)$, an action that has the structure of an electrodynamic Weber-like force with the following coefficients in first approach :

$$F = -\frac{G M_\odot}{r^2}\left(1 - 3\left(\frac{\dot{r}}{c}\right)^2 + 6\frac{r\ddot{r}}{c^2}\right)$$

The mathematical value of this force (for unit-mass of the object), produces an equation of motion that lightly perturbs the keplerian orbit inducing a precession with the same value as observed in Mercury`s orbit: ≈ 43" of arc per century. It is the same value obtained by General Relativity, although it is necessary to underline that however is the same precession, it is not the same general equation of motion. The precession for each orbit produced by the potential $U(r, Vr)$, is:

$$\Delta = \frac{6\pi G M_\odot}{c^2 p}$$

We can conclude that a newtonian potential perturbed by the transit action based upon the radial speed of an object, would produce an identical precession to the one observed for the planet Mercury and to any target with an elliptic trajectory. This is only true if the hypothetical propagation speed of the gravitational potential, is the same as the speed of light.

Let us examine the potential $U(r, Vr)$ from another alternative perspective:

$$U(r, Vr) = -\frac{G M_\odot}{r}\left(\frac{t_2}{t_1}\right)^2 = -\frac{G M_\odot}{r}\left(\frac{1}{1 - Vr/c}\right)^2 = -\frac{G M_\odot}{r}\left(1 + \frac{Vr}{c - Vr}\right)^2$$

$$\approx -\frac{G M_\odot}{r}\left(1 + \frac{Vr}{c}\right)^2$$

This last approach is perfectly acceptable because being the difference in order of $10^{-9}$, allows an easier analysis of the question. We will define $S = Vr/c$ as positive when (Vr) and (c) are in the same forward direction and negative in the opposite.

$$U(r, Vr) = U(r, S) = -\frac{G M_\odot}{r}(1+S)^2 = -\frac{G M_\odot}{r}(1 + S^2 + 2S)$$

The resultant potential is similar to the classic newtonian potential added with two perturbation terms, the first one proportional to $S^2$ and the second to $2S$.

The perturbance field $U_p(1 + S^2)$ was studied by Tisserand (1872) concluding that it produces a Weber force :

$$\mathbf{F_d} = -\frac{G M_\odot}{r^2}\left(1 - \left(\frac{\dot{r}}{c}\right)^2 + 2\frac{r\ddot{r}}{c^2}\right)$$

that becomes a precesión that is exactly 1/3 of the observation value:

$$\Delta_d = \frac{2\pi G M_\odot}{c^2 p}$$

Keeping in mind that the potentials U(r, Vr) and U(r, S) are equivalent, both must produce the same resultant precession; we will conclude that the precession linked with the perturbation term U$_p$(2S) is :

$$\Delta_L = \frac{4\pi G M_\odot}{c^2 p}$$

We will apply the Lagrangian equations to analyse what associated force is produced by the perturbation field U$_p$(2S):

$$F_L = \frac{d}{dt}\left(\frac{\partial U}{\partial \dot{r}}\right) - \frac{\partial U}{\partial r} \quad ; \text{ applying it to U}_p(2S):$$

$$\frac{\partial U}{\partial r} = -\frac{2 G M_\odot}{r^2} \cdot \frac{\dot{r}}{c} \quad ; \quad \frac{\partial U}{\partial \dot{r}} = \frac{2 G M_\odot}{r} \cdot \frac{1}{c} \quad ; \quad \frac{d}{dt}\left(\frac{\partial U}{\partial \dot{r}}\right) = -\frac{2 G M_\odot}{r^2} \cdot \frac{\dot{r}}{c}$$

The associated force is $F_L = 0$. This absence of effective force for a perturbation potential that depends lineally on the ratio between the radial and propagation speeds, can be surprising because the potential causes clear real effects which is the precession of the elliptic orbits.

In my opinion, the only real alternative with a null perturbating force, would be that the own newtonian acceleration, should produce a mechanical impulse due to the relative increase/decrease of the differential transit time inside the object. This mechanical momentum doesn't produce any added real force, ( it could have an apparent mathematical value), and is transmitted to the object modifying the angle of its speed with the tangent to the orbit. The mechanical impulse produces a differential turn on the trajectory and its consecutive amount along a complete orbit, reaches the observed final precession. It is necessary to underline that the differential kinematic precession -$d\delta$- produced by the mechanical impulse -$L_m$- associated with the potential U$_p$(2S), is twice as much the precession -$d\beta$- produced by the force $F_d$, a real dynamic force associated with the potential U$_p$(S$^2$). Another way of testing this relation between -$d\delta$- and -$d\beta$- is to analyse the kinematic and dynamic balance in any point of the orbit:

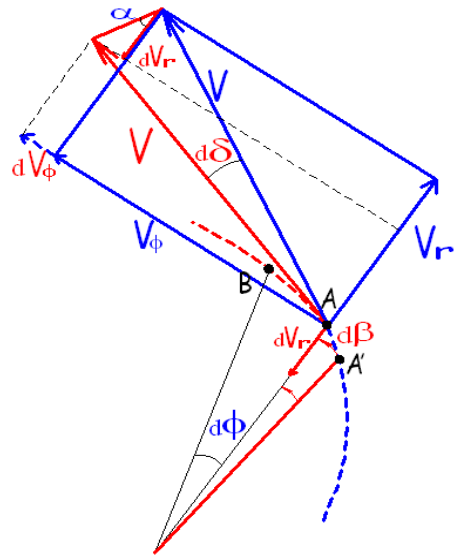

Fig –1

The radial differential speed caused by a mechanical momentum due to any perturbance acceleration ($a_p$) will be:

$$dV_r = a_p \cdot dt$$

As a consequence of $dVr$ presence, the tangent to the orbit will rotate an angle ($d\delta$) as result of the sum of the speeds.

$$dV_r = V \cdot d\delta \cdot Cos(\alpha); \quad d\delta = a_p / V_\varphi \cdot dt$$

The conservation of angular momentum ($h$), as it applies to any central force because such a force imposes no torque on the system, means :

$$h = V_\varphi \cdot r = \dot{\varphi} \cdot r^2$$
$$d\delta = r / h \cdot a_p \cdot dt = r^3 / h^2 \cdot a_p \cdot d\varphi$$

Besides the angle -$d\delta$-, another angle -$d\beta$- takes place also as a result of the balanced compensation of the new attraction force due to the added potential $U_p(S^2)$, that increases the newtonian potential in the upward branch of the ellipse and diminishes in the descending branch. The angle $d\beta$, doesn't means in itself an increment of the rotation speed; it is an angular deformation of the trajectory because the planet reacts to the perturbation force locating itself in the keplerian ellipse, in a position in which the centrifugal acceleration is equal to the newtonian added to the perturbation. This means that $a_p$ will be compensated in an "upstream" point of the canonical ellipse linked to an angle ($\varphi - d\beta$) "dragging" forward the keplerian ellipse.

The new centrifugal acceleration will be:

$$a_c = (\dot{\varphi} - \dot{d\beta})^2 \cdot r = (\dot{\varphi}^2 - 2\dot{\varphi}\,\dot{d\beta}) \cdot r ; \quad ( \gg \dot{d\beta}^2)$$

From the previous expression it is deduced that the centrifugal acceleration corresponding to $a_p$ is the second term $(2\dot{\varphi} \cdot \dot{d\beta}) \cdot r$

$$a_p = (2\dot{\varphi} \cdot \dot{d\beta}) \cdot r ; \quad \dot{d\beta} = d\beta / d\varphi \cdot \dot{\varphi}$$
$$d\beta = \tfrac{1}{2} \cdot r^3 / h^2 \cdot a_p \cdot d\varphi$$
$$d\beta = \tfrac{1}{2} \cdot d\delta$$

Therefore, the transit action produces a real force -$F_d$- and a mechanical momentum -$L_m$- whose combined action causes a precession ($d\beta + d\delta$) that is the same as the one observed for Mercury and also agrees with General Relativity elliptic orbits precession.

Along the descending branch of the ellipse, Mercury comes closer to the Sun with a radial speed opposite to the gravitational potential, therefore the final acceleration, will be something less compared with the classic gravitational one. The perturbing acceleration is directed outside the orbit, so the Planet will move outward in relation with the position it should occupy in the keplerian ellipse; this movement only takes place if the orbit rotates a forward angle ($d\delta + d\beta$). In both cases the precession takes place in the forward direction because in both branches of the ellipse, the balance point is located in an upstream position. Another form of understanding what happens is that in both branches, the keplerian ellipse is "dragged" forward as consequence of the turn ($d\delta + d\beta$) applied to the orbit of Mercury.

It is not a new axis-rotation added to the classic newtonian target-rotation in the keplerian orbit ; it is an interference of the orbit, a pulsation that produces a turn to the tangent of the ellipse, inward during the up-way branch and outward during the descending branch.

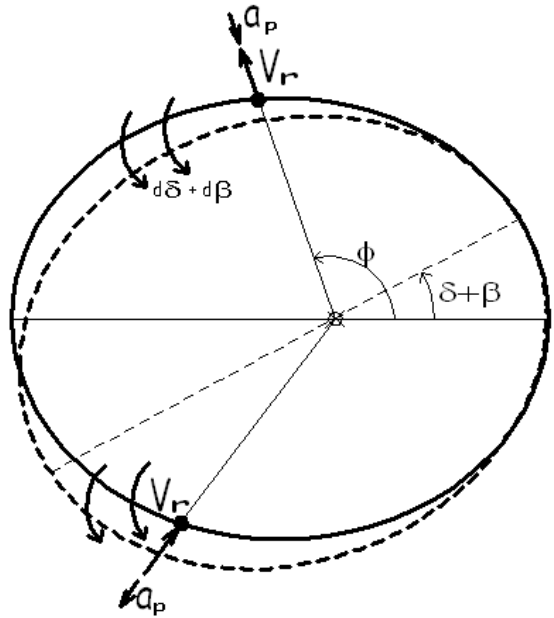

**Fig-2**

If it is really this way, the advance of the perihelion of Mercury would be the result of the hypothetical existence of a gravitational potential in the real universe of material objects; the energy transmitted to the target, is proportional to the square of the transit time. This value is obtained if propagation speed, is the same as speed of light, statement which is not clearly proven in spite of the results quoted by Kopeikin (2002). The aberration due to a gravitational potential of finite speed, is not significant in the solar system where the global barycentre stays inside the Sun. However, aberration happens in a double star system but also in that case, a balanced rotation is obtained although the gravitational acceleration is larger than the one that would fit with the classic law.

2.- **Pioneer 10/11 Anomalous Acceleration**.

The Pioneer 10/11 are spacecrafts whose mission was to explore the outer planets of the Solar System in particular Jupiter and Saturn. They travel with a nearly uniform speed of ≈12 Km/seg. Anderson et al, detected in 1998 a constant anomalous acceleration in both spacecrafts of $8,74 (\pm 1,33) \cdot 10^{-10}$ m/s$^2$ directed toward the Sun, between 20 and 60 AU.

If the gravitational potential $U(r,S)$ should exist with the characteristics previously defined, an anomalous acceleration directed toward the Sun would take place since the Pioneer10/11 travels with a considerable speed moving away from it. Therefore we will have a larger transit time between the gravitational potential and the spacecraft compared with a perfect circular orbit or a rest position.

We will apply $U(r,S)$, to the external navigation of the Pioneer 10/11 with an almost uniform relative speed of $Vr = 12.800$ m/s.

The anomalous acceleration produced by the " transit action" of the potential $U(r,S)$, should be a combination of ($F_d$) added to the mechanical momentum ($L_m$). The action taken place by $L_m$, would be the application result of the newtonian acceleration in an interval of time something larger compared with the transit time inside Pioneer if motionless or moving perpendicular to the Sun. The interference coefficient will be the ratio between the perturbed differential time ($t_2 - t_1$) and the newtonian time ($t_1$). The speed of Pioneer is nearly uniform, so we can consider $\gg \ddot{r}$. Being K the number of astronomical units (AU) equivalent to the distance between the ship and the Sun, we will have:

$$\frac{t_2 - t_1}{t_1} = S$$

$$a_p = - \frac{G M_\odot}{r^2} \left( S^2 + 2S \right) = - \frac{G M_\odot}{(K \cdot AU)^2} \left( 8{,}5386 \cdot 10^{-5} \right) = - \frac{1}{K^2} \cdot 5{,}054 \cdot 10^{-7} \text{ m/s}^2$$

For K=24 (AU), we obtain exactly the same anomalous acceleration directed toward the Sun as the Pioneer 10/11 have. For a distance of 10 AU, the anomalous acceleration should have reached values in order of $50{,}5 \cdot 10^{-10}$ m/s$^2$ although it has not probably been separated from the own classic gravitation because the flyby of Jupiter and Saturn located 5 and 10 AU. At a distance of 50 AU, the anomalous acceleration would reach $2{,}1 \cdot 10^{-10}$ m/s$^2$ although the speed should be fallen lightly. It is possible that this mechanical impulse, also affects the communication signal between spacecrafts and Earth, inducing the Doppler with an apparent acceleration of the Pioneer directed toward the Earth that would be added to the real previous one that is directed toward the Sun.

Although the potential $U(r,S)$ could explain at least partially the anomalous acceleration of aircrafts Pioneer 10/11, the most appropriate would be to carry out a new launching of another space ship. This mission is designed to determine the origin of the discovered anomaly, project that has been supported by the international scientific community.

### 3.- Deflection of Light in a Gravitational Field.

The deflection of light in a gravitational field, according to the principles of General Relativity, is equal to an angle of 1,75 seconds of arch, twice the value obtained by the classic newtonian mechanics. A perturbed gravitational potential similar to $U(r,S)$, should produce the relativistic deviation of light as the real force $F_d$ applied is:

$$F_d = - \frac{G M_\odot}{r^2} \left( 1 - \left( \frac{\dot{r}}{c} \right)^2 + 2 \frac{r \ddot{r}}{c^2} \right)$$

The coefficients of each term of the unitary force $F_d$, must have these exact values, just to produce the correct deflection of light and to be also a conservative energy system.

No conservative gravitational Weber-like force considered itself alone, can explain the precession of the perihelion of Mercury and the deviation of light at the same time.

However, the transit action of a potential $U(r,S)$, should produce a real Tisserand parameters acceleration -$F_d$- added with a mechanical impulse -$L_m$- that could make compatible both events if:

a) $F_d$ has indeed the correct coefficients linked with the deviation of light.
b) The kinematic effect produced by $L_m$ must not yield any deviation of the angle neither any perturbation to light photons besides a red or blue-shift detected by Doppler, or perhaps the light signal time-delay.

Anyway, this hypothetical conjecture is in front of a clear objection because as light is considered a target, -$V_r$- speed has the same magnitude as -$c$-, so there are not valid the approaches applied to objects with a "normal" speed as we have done with $U(r,S)$.

### 4.- Gravitational Field Statistic

As it was developed in 1993, if the gravitational potential is a suitable field in the physical-material universe, it cannot cover at the same time the whole entire space that occupies. Certainly its appearance of continuity and uniform density is extraordinary high due to the also extraordinary masses of the stars and source planets. Anyway, if its constitution is not absolutely continuous neither infinite, gravitation could be analyse by statistical procedures.

If we study the production of a gravitational potential from a focus, the law of the inverse square distance, could be obtained as the medium value of an aleatory force that verifies, a) the probability of the gravitational potential inside the stereoscopic angle of the target, b) an straight line trajectory inside the object and, c) the probability of concurrence of that specific trajectory in relation to the total mass of the attracted body.

If this potential $U(r,S)$, should exist hypothetically (although we don't have any real or experimental confirmation of it), the universal gravitation constant (G), would be in connection with the unitary production of momentum by two electrodynamic fields in contact ("transit action").

References :

*Construcción del Universo.* David Layzer. Scientific American Books, Inc

*Mexahnka.* Landau y Lifshitz. Kniga

*Principles of Cosmology and Gravitation.* M. Berry. Cambridge University Press

*Indication from Pioneer 10/11, Galileo and Ulysses Data of an apparent anomalous weak long-range acceleration.* J.D.Anderson et al.gr-qc/9808081 v2 Oct-1998

*Curso de Astronomía General.* Bakulin, Kononovich, Moroz. Editorial MIR

*The propagation of Gravity in Space and Time.* P.Gerber

*El orígen y desarrollo de la Relatividad.* J. Sanchez Ron. Alianza Editorial

*Relativity Theory: Its Origin and Impact on Modern Though.* A. Einstein et al. J. Wiley & Son.

*The Hidden Universe.* M Disney . J.M. Dent & Sons Ltd.

*A Brief History of Time.* S. Hawking. Bantam Books

*Was Einstein Right?.* C.M.Will. Basic Books Inc.

*Cosmology.* Scientific American

*On The Origin Of The Deflection Of Light* Jaume Gine : physics/0512121

*Weber-like interactions and energy conservation.* F. Bunchaft and S. Carneiro; gr-qc/9708047